\documentclass[twocolumn,aps,prl,showpacs,preprintnumbers,amsmath,amssymb,10pt]{revtex4-1}

\usepackage{graphicx}
\usepackage{dcolumn}
\usepackage{setspace}
\usepackage{bm}
\usepackage{multirow}

\begin{document}

\title{Disentangling the role of vibration, rotation, and neutron transfer in the fusion of neutron-rich mid-mass nuclei}

 \author{J. Vadas}
 \author{Varinderjit Singh}
 \author{B.~B. Wiggins}
 \author{J. Huston}
 \author{S. Hudan}
 \author{R.~T. deSouza}
 \email{desouza@indiana.edu}
 \affiliation{%
 Department of Chemistry and Center for Exploration of Energy and Matter, Indiana University\\
 2401 Milo B. Sampson Lane, Bloomington, Indiana 47408, USA}%

 \author{Z. Lin}
 \author{C.~J. Horowitz}
 \affiliation{%
 Department of Physics and Center for Exploration of Energy and Matter, Indiana University\\
 2401 Milo B. Sampson Lane, Bloomington, Indiana 47408, USA}%

 \author{A. Chbihi}
 \author{D. Ackermann}
 \affiliation{GANIL, 1 Blvd. Henri Becquerel, Caen, 14000, France}

 \author{M. Famiano}
 \affiliation{Department of Physics, Western Michigan University, Kalamazoo, Michigan 49008, USA}%

 \author{K.~W. Brown}
 \affiliation{National Superconducting Cyclotron Laboratory and Department
 of Physics and Astronomy, Michigan State University, East Lansing, Michigan 48824, USA}%

 \date{\today}

 \begin{abstract}
 We report the first measurement of the fusion excitation functions for $^{39,47}$K + $^{28}$Si 
 at near-barrier energies. Evaporation residues resulting from the fusion process were identified
 by direct measurement of their energy and time-of-flight with high geometric efficiency. 
 At the lowest incident energy, the cross-section measured for the neutron-rich $^{47}$K induced reaction is 
 $\sim$6 times larger than that of the $\beta$-stable system. 
 The experimental data are compared with both a dynamical deformation model and 
 coupled channels calculations (CCFULL). 
 \end{abstract}


 \maketitle

 The structure of neutron-rich nuclei and the reactions they undergo are presently a topic of considerable interest. 
 Not only are these nuclei relevant to r-process nucleosynthesis \cite{Goriely11} and the putative reactions 
 occurring in the outer crust of an accreting neutron star \cite{Horowitz08}, but they also provide the 
 opportunity to investigate weakly bound systems in which continuum effects are important \cite{Volya03}.
 For weakly bound neutron-rich nuclei near the limit of stability, 
 the weak binding of the valence neutrons allows these neutrons to readily couple to continuum states,
 modifying the properties of the bound states \cite{Volya03}. 
 With this increased coupling to continuum states, an increased collectivity is also expected. 
 The extent to which this increased collectivity manifests itself in nuclear reactions is an open question. 
 For example, one expects on general grounds that the enhanced polarizability associated with 
 exciting low-energy collective modes should result in an increased likelihood for fusion to occur. 
 Such general expectations are borne out by microscopic time-dependent Hartree-Fock (TDHF) calculations \cite{Umar12}. 
 Polarization of the nuclei during the collision can be envisioned as the prelude to neutron transfer. 
 While for stable beams, the impact of neutron transfer on fusion at near barrier energies has been shown to be significant 
 \cite{Zagrebaev12}, limited information exists for nuclei away from stability. 

 Radioactive beam facilities provide, 
 for the first time, the opportunity to systematically investigate the fusion of nuclei as they become 
 increasingly neutron-rich and the influence of coupling to the continuum increases. 
 In this Letter we present the first measurement of the fusion excitation functions for 
 $^{47}$K + $^{28}$Si and $^{39}$K + $^{28}$Si.  
 The role of specific collective modes for these systems is demonstrated and the framework for 
 future fusion studies is established.

 \begin{figure}[]
 \includegraphics[width=8.6cm]{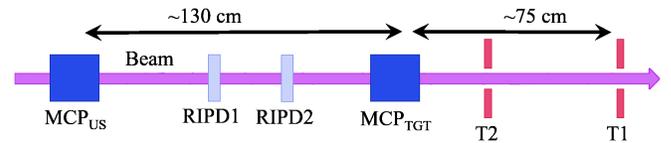}
 \caption{(Color online) Schematic of the experimental setup. 
 See text for details.
 }
 \label{fig:Setup}
 \end{figure}

 This experiment was performed using the ReA3 facility at 
 Michigan State University's National Superconducting Cyclotron Laboratory (NSCL).
 Ions of $^{39}$K from a source or $^{47}$K from a thermalized radioactive beam 
 were charge bred in an ion trap,  
 injected into the ReA3 80 MHz linac,
 and reaccelerated to energies between 2 and 3 MeV/A.
The experimental setup used to measure fusion of potassium ions with silicon nuclei is depicted in Fig.~\ref{fig:Setup}.
 The beam first passed through an upstream E$\times$B microchannel plate (MCP) detector, designated MCP$\mathrm{_{US}}$, 
 followed by another MCP detector in the target position (MCP$\mathrm{_{TGT}}$) 
 approximately 1.3 m downstream of the MCP$\mathrm{_{US}}$. 
 These detectors \cite{Steinbach14} provided a time-of-flight (TOF) measurement of the beam particles. 
 The 327 $\mu$g/cm$^{2}$ thick $^{28}$Si secondary emission foil of the MCP$\mathrm{_{TGT}}$ 
 served as the target for the experiment. 
 This time-of-flight measurement allowed rejection of beam particles 
 scattered or degraded prior to the target and provided a direct measure of the
 number of beam particles incident on the target.
 The intensity of the $^{39}$K beam on the target was 3 - 4.5 $\times$ 10$^4$ ions/s, and 
 that of the $^{47}$K beam was 1 - 2.5 $\times$ 10$^4$ ions/s.

 In order to identify contaminants in the $^{47}$K beam, two compact axial field ionization chambers \cite{Vadas16} 
 designated RIPD1 and RIPD2 were inserted in the beam path between the two MCP detectors.
 Particle identification was achieved by $\Delta$E-TOF, where the time-of-flight for each particle was 
 measured between the MCP detectors. 
 The energy distribution of incident $^{39}$K and $^{47}$K ions was measured by 
 periodically inserting a silicon surface barrier detector just upstream of the target. 
 The width, $\sigma$, of the energy distribution was $\sim$300 keV for $^{39}$K 
and $\sim$600 keV for $^{47}$K.

 Fusion of a $^{39}$K (or $^{47}$K) projectile nucleus with a $^{28}$Si target nucleus produces an excited 
 $^{67}$As ($^{75}$As) compound nucleus (CN). Near the 
 fusion barrier, the excitation energy of the CN 
is $\sim$40 MeV ($\sim$55 MeV). De-excitation of the CN via evaporation of light particles 
 imparts transverse momentum to the evaporation residue (ER), allowing its detection in 
 the annular silicon detectors designated T1 and T2 which 
 subtend the angles 1.0$^\circ$ $\le$ $\theta_{lab}$ $\le$ 7.3$^\circ$.

 \begin{figure}[]
 \includegraphics[width=8.6cm]{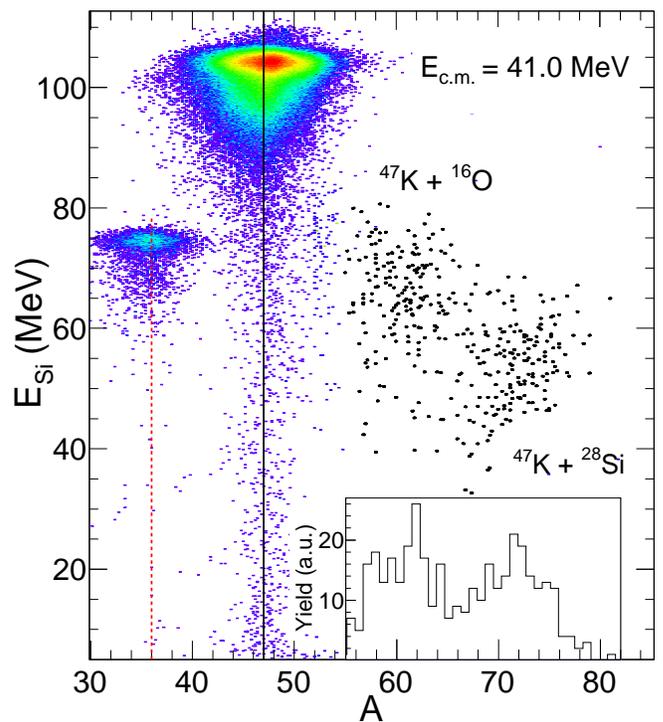}
 \caption{(Color online) Energy versus mass number for reaction products 
 with 2.4$^\circ$ $\le$ $\theta_{lab}$ $\le$ 7.3$^\circ$. For reference, 
 a solid (black) line and dashed (red) line are shown at A=47 and A=36, respectively. 
 Evaporation residues are shown in bold.
 Inset: Mass distribution of evaporation residues.
 }
 \label{fig:EvsA}
 \end{figure}

 To distinguish ERs from scattered beam, reaction products detected in the silicon detectors are identified by their mass using
 the energy vs. time-of-flight (E-TOF) technique \cite{Steinbach14,SteinbachPRC14,Singh17},
with the time-of-flight measured between the MCP$\mathrm{_{TGT}}$ 
 and the silicon detectors \cite{deSouza11}. 
 This mass identification for the $^{47}$K beam is presented in Fig.~\ref{fig:EvsA}. 
 The intense peak evident at E$\mathrm{_{Si}}$=105 MeV and A$\approx$47 
 corresponds to elastically scattered $^{47}$K particles, 
 with the points extending down in energy along the solid A=47 line
 corresponding to scattered $^{47}$K particles. 
 Elastically scattered $^{36}$Ar, a beam contaminant, is evident at 
 E$\mathrm{_{Si}}$=75 MeV centered on A=36, and accounts for $<$1$\%$ of the 
 incident beam particles. Two distinct islands cleanly separated from the scattered beam particles are apparent at high mass (shown in bold). 
 The inset shows the mass distribution of these islands, where a clear separation between the two islands 
 is observed at A$\approx$66. The lower mass island around A=60 corresponds to 
 evaporation residues of compound nuclei formed following fusion of $^{47}$K projectile nuclei 
 with $^{16}$O nuclei present in an oxide layer on the target foil (A$\mathrm{_{CN}}$=63).
 The higher mass island peaking at A$\approx$72 is populated by ERs following fusion of $^{47}$K + $^{28}$Si, which is consistent with the prediction of a fusion-evaporation code, EVAPOR \cite{evapOR}. 
ERs from the $^{28}$Si target were selected using the two-dimensional E$_{Si}$ {\it vs} A spectrum.

 The measured yield of evaporation residues (N$\mathrm{_{ER}}$) was used to calculate 
 the fusion cross-section $\sigma\mathrm{_{fusion}}$  using 
 $\sigma\mathrm{_{fusion}}$ = N$\mathrm{_{ER}}$ / ($\epsilon\mathrm{_{ER}}$ $\times$ N$\mathrm{_{Beam}}$ $\times$ t), 
 where $\epsilon\mathrm{_{ER}}$ is the detection efficiency, 
 N$\mathrm{_{Beam}}$ is the number of beam particles incident on the target and {\it t} is the target thickness. 
 N$\mathrm{_{Beam}}$ was determined using the time-of-flight between the
 MCP detectors and particle identification in the $\Delta$E-TOF spectrum. 
The thickness of the target
 was gauged using $\alpha$ particles from sources \cite{SRIM}. 
 After accounting for the oxide layer, the $^{28}$Si thickness was determined.
 The detection efficiency $\epsilon\mathrm{_{ER}}$ was calculated by using EVAPOR \cite{evapOR} 
 and the geometric acceptance of the silicon detectors. For both systems, 
 the combined geometric efficiency of T1 and T2, for all incident energies measured, was $\sim$80$\%$.

 \begin{figure}[]
 \includegraphics[width=8.6cm]{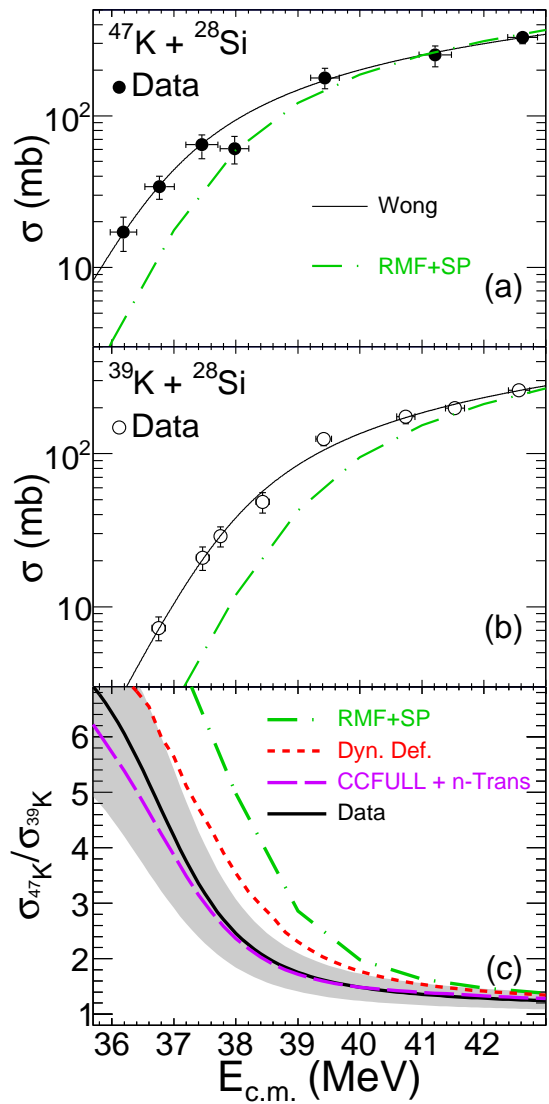}
 \caption{(Color online) 
 Panel (a):  
 The fusion excitation function for $^{47}$K + $^{28}$Si.  
 Solid (black) line corresponds to a fit of the experimental data as described in the text. 
 Dash-dot (green) line corresponds to calculated fusion cross-sections using the Sao Paulo model \cite{Gasques04}
 with density distributions calculated within a relativistic mean field framework \cite{Ring96,Serot86}.
 Panel (b):
 Same as above for $^{39}$K projectile ions.
 Panel (c):
 The relative cross-section, $\sigma$($^{47}$K)/$\sigma$($^{39}$K)
 is depicted as a solid (black) line, corresponding to the ratio of the Wong fits of the experimental data.
 The shaded band represents the measurement uncertainty. The dashed (colored) lines correspond to model calculations
 described in the text.
 }
 \label{fig:XSectStatic}
 \end{figure}

 The fusion cross-sections as a function of incident energy for $^{39}$K + $^{28}$Si (open circles) 
 and $^{47}$K + $^{28}$Si (closed circles) are shown in Fig.~\ref{fig:XSectStatic}. Both fusion excitation functions exhibit the
 general trend expected for a barrier-driven process. With decreasing incident energy, the fusion cross-section decreases 
 slowly for energies above the barrier then drops dramatically at and below the barrier. To facilitate 
 comparison of the two systems, the measured fusion excitation functions were parameterized using a functional form 
 that describes the penetration of an inverted parabolic barrier (Wong formula) \cite{Wong73}.
 The fits of the $^{39}$K and $^{47}$K data are shown in Fig.~\ref{fig:XSectStatic} as the 
 solid (black) lines. With the exception of the $^{47}$K cross-section measured at E$\mathrm{_{c.m.}}$=38 MeV 
 the excitation functions for both systems are reasonably well described by this parameterization.

 The relative cross-section 
 $\sigma$($^{47}$K)/$\sigma$($^{39}$K) as a function 
 of incident energy is shown in Fig.~\ref{fig:XSectStatic}(c). 
 For energies above the barrier, the ratio is 
 essentially flat with a value of $\sim$1.2, but as E$\mathrm{_{c.m.}}$ decreases below the barrier, 
 the ratio rapidly increases to a factor of $\sim$6 at the lowest measured energy.

 In order to better understand the extent to which the observed fusion cross-sections are
 due to the nuclear size, structure, or dynamics, 
 we have calculated the fusion
 of potassium isotopes with $^{28}$Si nuclei with different models. The simplest model 
 utilized is a Sao Paulo (SP) model \cite{Gasques04},
 which allows one to assess the changes in the fusion cross-section due solely to the changes in the 
 density distributions of the nuclei. These density distributions
 have been calculated within a relativistic mean field (RMF) model \cite{Ring96,Serot86},
and were utilized 
 in a folding potential to predict the fusion cross-sections.
 As the density distributions are spherically symmetric, initial deformation of the projectile and target nuclei
 are ignored in this approach.
 The cross-section predicted from the RMF+SP model is depicted in Fig.~\ref{fig:XSectStatic} as the dash-dot (green) line. 
 While this static model provides reasonable agreement at above-barrier energies, it significantly underpredicts
 the measured cross-sections at energies near and below the barrier, indicating that the size of the colliding nuclei 
 alone is insufficient to explain the observed fusion cross-sections.

 \begin{figure}[]
 \includegraphics[width=8.6cm]{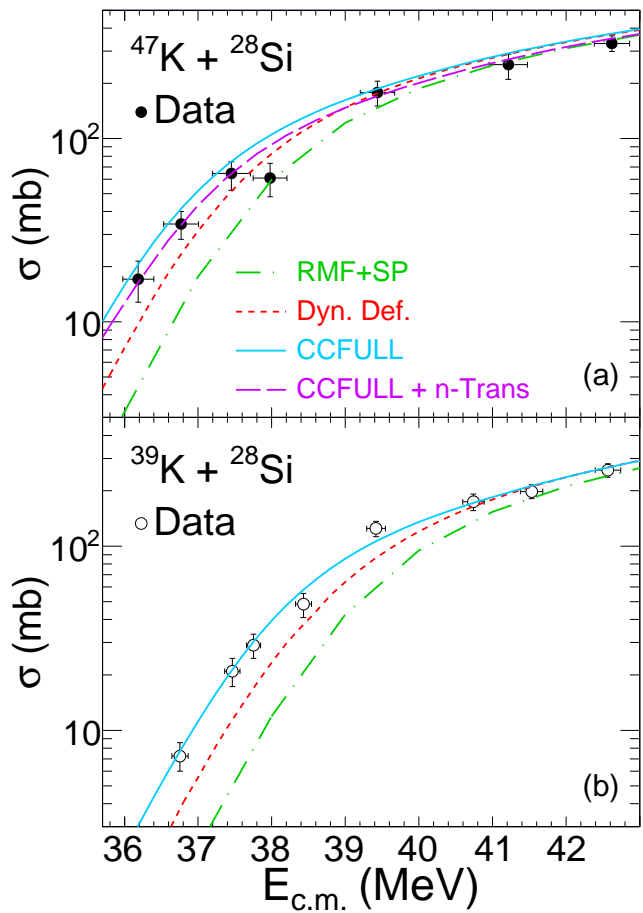}
 \caption{(Color online) 
 Panel (a):  
 Experimental data for $^{47}$K + $^{28}$Si are represented by symbols. 
 Solid and dashed (colored) lines correspond to different models described in the text. 
 Panel (b): 
 Same as above for $^{39}$K projectile ions.
 }
 \label{fig:XSectDynamic}
 \end{figure}

 \begin{table}[h]
 \centering
 \caption{\label{tab:5/tc}Woods-Saxon potential parameters for the measured systems.}
 \begin{tabular}{|c|c|c|c|}
 \hline
                     & $V_0$ (MeV) & $r_0$ (fm) & $a$ (fm) \\
 \hline
 \rule{0pt}{2.5ex} 
 $^{39}$K~+~$^{28}$Si & -55.03 & 1.16 & 0.612\\

 $^{47}$K~+~$^{28}$Si & -55.97 & 1.16 & 0.622\\
 \hline
 \end{tabular}
 \label{table:WSParams}
 \end{table}

 For mid-mass stable nuclei, the role of dynamics (collective modes) in describing the fusion cross-section 
 in the near-barrier regime is well established \cite{Umar07,Montagnoli13}.
 While a microscopic description of the fusion process with a model such as density constrained TDHF 
 \cite{Umar06d,deSouza13} is desirable, 
 calculation of fusion for the odd-A nuclei in these systems is particularly challenging. 
 As pairing significantly impacts the fusion cross-section \cite{SteinbachPRC14}, a correct treatment of the 
 unpaired nucleon is essential and presently beyond the scope of the model. 
 We therefore elected to explore the role of dynamics using two different approaches:
 a dynamical deformation model \cite{Zagrebaev01} and coupled channels calculations \cite{Beckerman88}.

 Within the dynamical deformation model, as the two nuclei approach, 
 coupling to collective degrees of freedom changes the potential barrier, resulting in 
 a distribution of barriers \cite{Rowley91} over the collective mode.
 The total penetration probability of the nuclear wavefunctions is averaged over 
 the barrier distribution with respect to deformation of the system \cite{Zagrebaev01}. 
 The Coulomb part of the potential is given by Coulomb's law, modified at small internuclear separation to 
 account for overlap.
 The nuclear part of the potential utilized a Woods-Saxon form, 
 with the depth V$_{0}$, radius parameter r$_{0}$, and diffuseness parameter {\it a}. 
 The values of these parameters were chosen to reflect the same 
 effective potential as that of the RMF+SP calculations, and are presented in Table~\ref{table:WSParams}.

 As evident in Fig.~\ref{fig:XSectDynamic} the dynamical deformation calculations indicated by the
 dotted (red) line are closer to the measured cross-sections for both systems than the RMF+SP calculations, 
 but still underpredict the measured values at lower energies. This indicates that coupling to surface vibrations
 is insufficient to describe the fusion cross-sections for these systems. As inclusion of surface vibrations in this model corresponds to the continuous limit of coupling to discrete vibrational states,
 this result indicates that coupling to vibrational states alone cannot describe the experimental fusion data.

 The inclusion of coupling to collective states in the potassium and silicon nuclei can also be treated 
 in the context of coupled channels calculations, using the code CCFULL \cite{Hagino99}. 
 For these calculations, coupling to the 1/2+ ground state and the 3/2+ and 5/2+ 
 excited states of $^{47}$K were included, and considered to be members of a rotational band. For $^{39}$K, 
 the 3/2+ ground state and excited 5/2+ and 7/2+ levels were included.
 Coupling in the $^{28}$Si target nucleus included the 2+ and 4+ levels of the rotational band
 built on the 0+ ground state.
 The results of these calculations are shown in Fig.~\ref{fig:XSectDynamic} as the solid (blue) line.
 While in the case of $^{39}$K + $^{28}$Si the CCFULL calculations provide a good description of the
 experimental cross-sections, for the $^{47}$K induced reaction the model slightly overpredicts the data. 
 Although neutron transfer plays no role in the case of $^{39}$K due to the negative Q-value, in the case of
 $^{47}$K, Q$_{2n}$ = 3.844 MeV, suggesting that neutron transfer may play a role.
 We therefore included neutron transfer channels in the CCFULL calculations of the $^{47}$K induced reaction.
 Inclusion of a weak coupling (F$_{t}$ = 0.25) to the neutron transfer channels provides the best description 
 of the experimental data, 
 as shown by the dashed (purple) line in Fig.~\ref{fig:XSectDynamic}(a).
 Increasing the coupling strength for neutron transfer above F$_{t}$ = 0.35 significantly underpredicts 
 the measured cross-sections, especially above the barrier.

 It is additionally instructive to compare the relative cross-section, $\sigma$($^{47}$K)/$\sigma$($^{39}$K),
 predicted by the various models with the experimental data as presented in Fig.~\ref{fig:XSectStatic}(c).
 At the highest energies measured, all of the model calculations converge and are in good agreement with the experimental data.
 This result is unsurprising, as the cross-section at above-barrier energies are dictated by 
 the geometric cross-section.
 The RMF+SP model exhibits a more rapid increase with decreasing incident energy than the data.
 While the dynamical deformation model manifests a smaller relative cross-section than the RMF+SP, it
 still predicts a larger relative cross-section than the data.
 As expected from their agreement with the excitation functions, 
 the CCFULL calculations including neutron transfer provide a good description of the measured relative cross-section.
 A small underprediction, within the measurement uncertainties, is observed at the lowest measured energies.

 The comparison of the relative cross-section predicted by the different models with the experimental data 
 can be understood by considering the cross-section as having both a geometric component as well as a dynamical component
 ($\sigma_{fusion}$ = $\sigma_{geometric}$ + $\sigma_{dynamic}$).
 If the dynamical contribution is dominated by the geometric component ($\sigma_{fusion}$ $\approx$ $\sigma_{geometric}$), 
 the experimental data should follow the 
 predictions of the RMF+SP model. Since the experimental data lies below the RMF+SP predictions, this indicates that 
 not only do dynamics play a role, as was evident in the fusion excitation functions, 
 but dynamics for $^{39}$K plays a larger role relative to the geometric cross-section 
 than in the case of $^{47}$K. 
 The relative cross-section predicted by the CCFULL calculations (with neutron transfer),
 while in good agreement with the experimental data within the statistical uncertainties, 
 exhibits a slight underprediction at the lowest energies.
 The relative fusion cross-section is therefore useful in comparing the 
 neutron-rich system with the $\beta$-stable system as 
 it cancels the average incident energy dependence associated with the barrier penetration process, providing higher sensitivity.

 Using a reaccelerated radioactive beam,
 the fusion excitation function for $^{47}$K + $^{28}$Si has been measured 
 for the first time, and is compared to the fusion excitation function for $^{39}$K + $^{28}$Si.
 At near and sub-barrier energies, a dramatic, six-fold increase in the fusion cross-section 
 is observed for the neutron-rich system
 relative to the $\beta$-stable system.
 Within a static approach with only spherically symmetric density distributions for the colliding nuclei,
 the fusion excitation function for both systems falls more steeply with decreasing incident energy
 than is experimentally observed, a result attributable 
to the role of structure and dynamics in the fusion process.
 Calculations with a dynamical model which include coupling to surface vibrations provide 
 a better description of the excitation functions, but still underpredict the data.
 By utilizing a coupled channels approach with coupling to measured rotational states, 
 a good description of the fusion excitation functions is achieved.
 For the neutron-rich projectile, $^{47}$K, a weak coupling to neutron transfer channels provides 
 a slightly better description.
 Comparison of the relative cross-sections predicted by the RMF+SP and CCFULL models with the experimental data
 indicates that the contribution of the dynamics to the fusion cross-section for the $^{39}$K induced reaction 
 is larger than in the case of $^{47}$K relative to the geometric cross-section.
 Comparison of the cross-sections for the neutron-rich systems with the relevant $\beta$-stable system 
 allows one to evaluate the extent to which the contribution of dynamics relative to the geometric cross-section
 evolves with neutron number.  

 As demonstrated by this work, measurement of high quality near-barrier fusion excitation functions
 with radioactive beams allows investigation of the role of 
 vibration, rotation, and neutron transfer in fusion.
 Measuring the fusion excitation function for an isotopic chain of neutron-rich
 light and mid-mass nuclei is therefore a powerful tool to examine the evolution of collective modes 
 and coupling to the continuum with increasing neutron number.
 A new generation of radioactive beam facilities, both in existence as well as on the horizon, 
 now makes such measurements possible for the first time.

\begin{acknowledgments}
 We thank the staff at NSCL, Michigan State University, and in particular those at the ReA3 facility 
 for providing the high quality beams that made this experiment possible. 
 The high quality $^{28}$Si targets, provided by M. Loriggiola, Legnaro National Laboratory, are deeply appreciated. 
 This work was supported by the 
 U.S. Department of Energy under Grant Nos. DE-FG02-88ER-40404 (Indiana University), DE-FG02-87ER-40365 (Indiana University Nuclear Theory), 
 DE-SC0008808 (NUCLEI SciDAC Collaboration), DE-NA0002923 (Michigan State University),
 and the National Science Foundation under Grant Nos. PHY-1565546 (Michigan State University) and PHY-1712832 (Western Michigan University).
 J.V. acknowledges the support of a NSF Graduate Research Fellowship under Grant No. 1342962.
One of us (D.A.) is supported by the European Commission in the framework of the CEA-EUROTALENT program.
\end{acknowledgments}

%

\end{document}